\documentclass[12pt,preprint]{aastex}  % one-column, single-spaced style
\newcommand{\HII}{\mbox{H\thinspace{\sc ii}}} % 

\shorttitle{The Magnetized Environment in W3(H$_2$O)} %
\shortauthors{Chen et al.} %

\begin{document} %
%%%%  Title  %%%%
\title{
The Magnetized Environment of the W3(H$_2$O) Protostars 
} %

%%%%  Authors  %%%%
\author{Huei-Ru Chen\altaffilmark{1,2}, Ramprasad Rao\altaffilmark{2}, David J. Wilner\altaffilmark{3}, and Sheng-Yuan Liu\altaffilmark{2}} %
%\affil{Institute of Astronomy \& Department of Physics, National Tsing Hua University, Hsinchu 30013, Taiwan} %
%\and %
%\author{other authors \altaffilmark{3}} %

%%%%  Affiliations  %%%%
\altaffiltext{1}{Institute of Astronomy and Department of Physics, National Tsing Hua University, Hsinchu, Taiwan; hchen@phys.nthu.edu.tw.} %
\altaffiltext{2}{Institute of Astronomy and Astrophysics, Academia Sinica, Taipei, Taiwan.} %
\altaffiltext{3}{Harvard-Smithsonian Center for Astrophysics, Cambridge, MA.} %

%%%%  Abstract  %%%%
\begin{abstract} %  
We present the first interferometric polarization map of the W3(OH) massive star-forming region observed with the Submillimeter Array (SMA) at $878 \; \mu\mathrm{m}$ with an angular resolution of 1\farcs5 (about $3 \times 10^3 \; \mathrm{AU}$).  
Polarization is detected in the W3(H$_2$O) hot core, an extended emission structure in the north-west of W3(H$_2$O), and part of the W3(OH) ultracompact \HII\ region.    
The W3(H$_2$O) hot core is known to be associated with a synchrotron jet along the east-west direction.  
In this core, the inferred magnetic field orientation is well aligned with the synchrotron jet and close to the plane of sky.  
Using the Chandrasekhar-Fermi method with the observed dispersion in polarization angle, we estimate a plane-of-sky magnetic field strength of $17.0 \; \mathrm{mG}$.  
Combined with water maser Zeeman measurements, the total magnetic field strength is estimated to be $17.1 \; \mathrm{mG}$, comparable to the field strength estimated from the synchrotron model.  
The magnetic field energy dominates over turbulence in this core.  
In addition, the depolarization effect is discerned in both SMA and JCMT measurements.  
Despite the great difference in angular resolutions and map extents, the polarization percentage shows a similar power-law dependence with the beam averaged column density.  
We suggest that the column density may be an important factor to consider when interpreting the depolarization effect.  
\end{abstract} %

%%%% Keywords  %%%%

%%%% Introduction  %%%%
\section{Introduction} %
%% Polarization emission
%\key{Implication of polarization observations.  Link to B fields.} %

One of the outstanding problems in star formation concerns the influence of magnetic fields to the dynamics and timescales \markcite{Shu1999CreteII,McKee2007ARA&A}(Shu {et~al.} 1999; McKee \& Ostriker 2007). 
The relative importance between magnetic fields and turbulence affects cloud evolution and cluster formation \markcite{McKee2007ARA&A}(McKee \& Ostriker 2007).  
Ambipolar diffusion has been proposed to regulate core collapse \markcite{Shu:1987dp}(Shu, Adams, \& Lizano 1987). 
In accretion flows, magnetic effects such as magnetic tension, braking, and reconnection, or non-ideal magnetohydrodynamic (MHD) effects, can drive the evolution of the formation process \markcite{Shu:2004ed,Li:2011ds}(Shu, Li, \& Allen 2004; Li, McKee, \& Klein 2011).  
The residual magnetic fields in the circumstellar environs further assist to collimate supersonic outflows \markcite{Shu:2000vf,Shang2007PPV}(Shu {et~al.} 2000; Shang, Li, \& Hirano 2007). 
Intringingly, radio emission with negative spectral index has been reported in a few protostellar jets, e.g. HH80-81 \markcite{Marti:1993ek,CarrascoGonzalez:2010dc}(Mart{\'\i}, Rodriguez, \&  Reipurth 1993; Carrasco-Gonzalez {et~al.} 2010), Cepheus~A \markcite{Garay1996ApJ459}(Garay {et~al.} 1996), $\mathrm{W3(H_2O)}$ \markcite{Reid:1995ib,Wilner1999ApJ513}(Reid {et~al.} 1995; Wilner, Reid, \& Menten 1999). 
Such non-thermal emission resembles jets driven by active galactic nuclei and suggests a connection with synchrotron jets.    

It has become increasingly clear that dust grains are aligned by magnetic fields not only in the diffuse interstellar medium but also in dense cloud cores or circumstellar disks \markcite{Myers:1991ky,WardThompson:2000fd,Lazarian:1997ht,Matthews:2009cj,Dotson:2000hm,Dotson:2010je}(Myers \& Goodman 1991; Ward-Thompson {et~al.} 2000; Lazarian, Goodman, \& Myers 1997; Matthews {et~al.} 2009; Dotson {et~al.} 2000, 2010).  
Although grain alignment mechanisms may be less assured, polarization remains one of the most  informative techniques to trace magnetic fields \markcite{Lazarian:2007do}(Lazarian 2007).
Complimentary to polarization from selective extinction of background starlight, polarized emission arising from magnetically aligned dust grains in the millimeter or submillimeter wavebands is often used to map the two-dimensional morphology of the three-dimensional (3D) magnetic fields  projected on the plane of sky, $B_\bot$.  
Depending on the type of polarization, the observed orientation is supposed to be either parallel to $B_\bot$ if observed in dust absorption against background sources or perpendicular to $B_\bot$ if observed in dust thermal emission \markcite{Hildebrand:1988tt}(see the review by Hildebrand 1988). 

The bright ultracompact (UC) \HII\ region, W3(OH), marks a nearby \markcite{Hachisuka:2006dz}($d = 2.04 \; \mathrm{kpc}$; Hachisuka {et~al.} 2006), well-studied massive star-forming region in the course of developing an OB stellar group.  
At least seven \HII\ regions at different evolutionary stages were found within 30\arcsec\ of W3(OH) \markcite{Harten:1976vn}(Harten 1976).  
The most recent star formation takes place in a luminous ($L \sim 10^4 \, L_\odot$) hot core, $\mathrm{W3(H_2O)}$ (a.k.a the Turner-Welch object), which is associated with water masers and appears as a warm, dense molecular clump about 6\arcsec\ east to W3(OH) \markcite{Turner:1984gl}(Turner \& Welch 1984).  
Millimeter and submillimeter interferometric studies resolved the dust emission into two components, A and C, and suggest an accreting proto-binary system \markcite{Wyrowski:1999cu,Chen2006ApJ639,Zapata:2011dy}(Wyrowski {et~al.} 1999; Chen {et~al.} 2006; Zapata {et~al.} 2011).  
Proper motions of water masers trace a bipolar outflow centered at $\mathrm{W3(H_2O)A}$ \markcite{Alcolea:1993gd,Hachisuka:2006dz}(Alcolea {et~al.} 1993; Hachisuka {et~al.} 2006), which also drives a synchrotron jet along the east-west direction \markcite{Reid:1995ib,Wilner1999ApJ513}(Reid {et~al.} 1995; Wilner {et~al.} 1999).
Here we perform interferometric polarization observations toward the W3(OH) massive star-forming region to map the morphology of the projected magnetic fields in this region.  
Our study is enabled by the Submillimeter Array\footnotemark[4] (SMA) with its polarimeter system  \markcite{Marrone2006ApJ640,Marrone:2008kl}(Marrone {et~al.} 2006; Marrone \& Rao 2008). 

\footnotetext[4]{The Submillimeter Array is a joint project between the Smithsonian Astrophysical Observatory and the Academia Sinica Institute of Astronomy and Astrophysics, and is funded by the Smithsonian Institution and the Academia Sinica.}  

%%%% Observations & Data Reduction %%%%
\section{Observations and Data Reduction \label{sec_Observations}} %
Observations using quarter-wave plates installed on the SMA were carried out with seven antennas in the extended configuration on 2007 August~10 and with six antennas in the compact configuration on 2008 September~20.    
The phase tracking center is at $(\alpha, \delta)$(J2000) = $\mathrm{(2^h27^m3.870^s,+61^\circ52^\prime24.60^{\prime\prime})}$.    
The system temperature varied from 150 to 580~K during the extended array observation and from 110 to 400~K during the compact array observation.  
The correlator was set to have a spectral resolution of $0.8125 \; \mathrm{MHz}$, corresponding to $0.7~\mathrm{km \, s^{-1}}$, and 2~GHz bandwidth in each sideband.
%The projected baselines compiled from all observations ranged from $8 - 207\; \mathrm{k\lambda}$, which are insensitive to structures larger than 11\arcsec\ \markcite{Wilner1994ApJ427}(Wilner \& Welch 1994). 
The projected baselines have a range of $8 - 207\; \mathrm{k\lambda}$, which is insensitive to structures larger than 11\arcsec\ \markcite{Wilner1994ApJ427}(Wilner \& Welch 1994). 
The full-width at half power (FWHP) of the primary beam is roughly 37\arcsec.  

%The observing loop comprised scans of 3C84 and W3(OH) was repeated every 27 minutes.

Data inspection, calibration, and imaging were performed with the MIRIAD package. 
The instrumental polarization response, i.e. ``leakage,'' for each antenna was calibrated by observing strong quasars, 3c111 and 3c454.3, over a wide hour angle range for a good parallactic angle coverage.   
The bandpass calibration was performed with 3c454.3.  
Ganymede and Callisto were used to set the flux density scale, which is estimated to be accurate within 15\%.
Regularly interleaved observations of the nearby quasar 3c84 were conducted to monitor the atmospheric variations.  
Atmospheric complex gains were first solved with respect to 3c84 to make initial  Stokes $I$ images for individual tracks.  
The gain solutions were further improved by one iteration of phase-only self-calibration using the initial Stokes $I$ images as models. 
To image continuum emission, care has been taken to remove channels with line emission.  
%Both sidebands, each of 2 GHz bandwidth, were used to generate line-free continuum maps in a multi-frequency synthesis, resulting a frequency of $341.5 \; \mathrm{GHz}$, equal to the average of the upper and lower sideband frequencies. 
Both sidebands were used to generate line-free continuum maps in a multi-frequency synthesis with an effective central frequency of $341.5 \; \mathrm{GHz}$.  

To obtain the best signal-to-noise ratio, maps of Stokes $I$ (total intensity), and Stokes $Q$ and $U$ (linear polarization) were made with visibilities weighted by the associated system temperature using natural weighting.  
% uniform weighting results only measurements in W3(H2O)
This results in a synthesized beam of $1\farcs4 \times 1\farcs5$ with position angle (P.A.) of $-14^\circ$. 
For easy comprehension and convenience of subsequent Nyquist sampling, we smooth all the maps with a circular Gaussian beam of $1\farcs5$.   
The resultant maps have an rms noise level of $\sigma_I = 27 \; \mathrm{mJy \, beam^{-1}}$ for the Stokes $I$ map and $\sigma_{QU} = 3 \; \mathrm{mJy \, beam^{-1}}$ for maps of Stokes $Q$ and $U$. 
The dynamic range of the Stokes $I$ map is limited at a maximum intensity value of $120 \, \sigma_I$ due to the $u$-$v$ coverage and calibration accuracy.    
Once Stokes $I$, $Q$, and $U$ are known, the polarization percentage, $p$, and the polarization position angle, $\theta$, can be deduced \markcite{Matthews:2000dg}(Matthews \& Wilson 2000).
The raw polarized intensity, $I_{p,\mathrm{raw}} = (Q^2+U^2)^{1/2}$, is necessarily positive and biased by the rms noise of Stokes $Q$ and $U$, $\sigma_{QU}$.  
Hence, a debiased polarized intensity is used, $I_p = (I_{p,\mathrm{raw}}^2 - \sigma_{QU}^2)^{1/2}$ \markcite{Leahy:1989vm,Vaillancourt:2006bf}(Leahy 1989; Vaillancourt 2006) when computing the polarization percentage, $p = I_p/I$.  
The polarization position angle is determined by $\theta_\mathrm{obs} = (1/2) \, \tan^{-1} (U/Q)$, with an uncertainty of $\sigma_{\theta,\mathrm{obs}} = (1/2)(\sigma_{QU}/I_p)(180^\circ/\pi)$.  
Note that in linear polarization measurements, values of $\theta_\mathrm{obs}$ different by $180^\circ$ are identical.  
After generating maps of Stokes $I$, $Q$, and $U$, we image the distribution of polarization, including $I_p$, $p$, and $\theta_\mathrm{obs}$, for regions where $I_p \ge 3 \, \sigma_{QU}$ and $I \ge 3 \, \sigma_I$.  
Table~\ref{tpol} lists the Nyquist sampling of the polarization map with a grid step of 0\farcs75.  
 
%%%% Results & Discussion %%%%
\section{Results and Discussion} % 
\begin{figure} %
\epsscale{1.0} %
\plotone{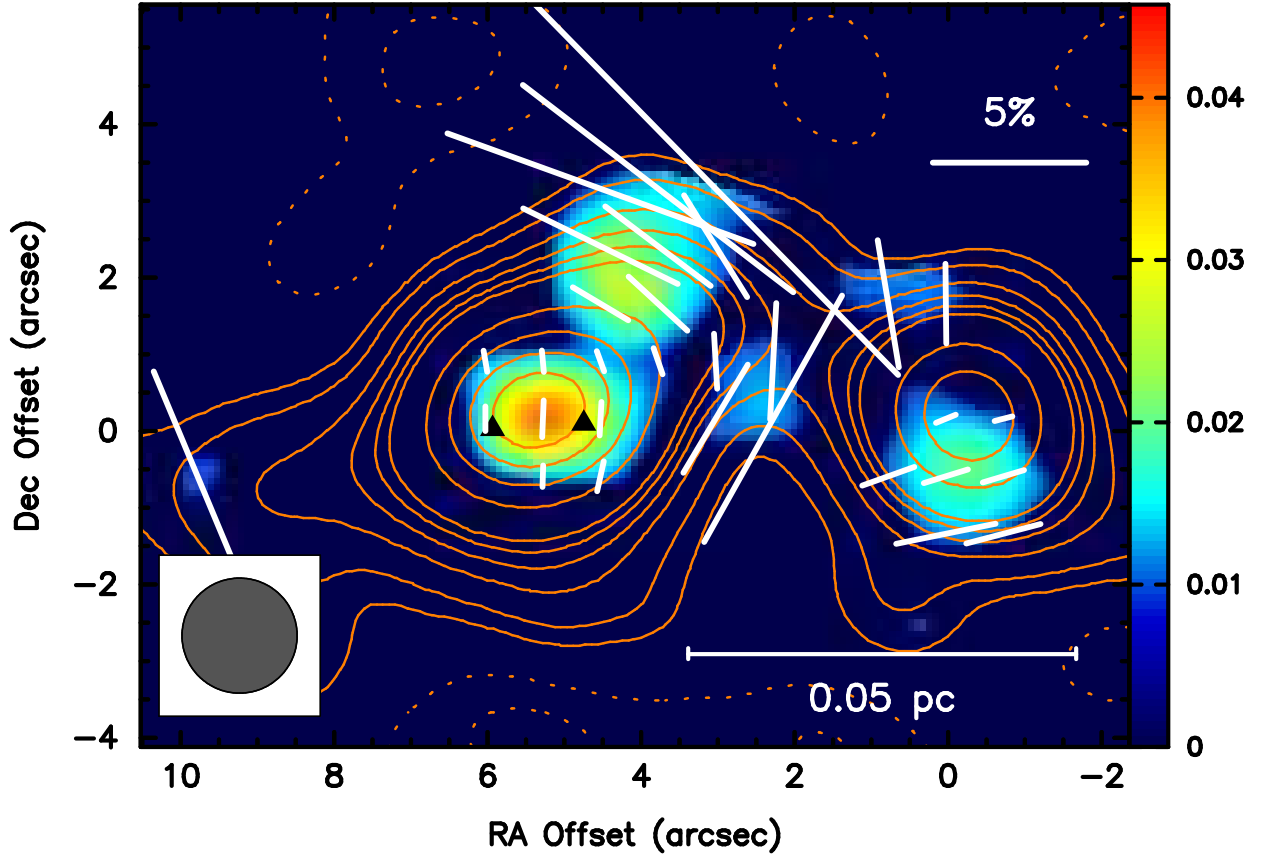} %
\caption{Polarization map of the W3(OH) region at $878 \; \mu\mathrm{m}$ with the total intensity (Stokes $I$) map (contours) overlaid on the linearly polarized intensity, $I_p$, map (colors).  
Line segments give polarization orientation with their lengths proportional to the polarization percentage and are plotted where $I_p \ge 3 \, \sigma_{QU}$ and $I \ge 3 \, \sigma_I$.   
The rms value is $\sigma_I = 27 \; \mathrm{mJy \, beam^{-1}}$ for the Stokes $I$ map and $\sigma_{QU} = 3 \; \mathrm{mJy \, beam^{-1}}$ for maps of Stokes $Q$ and $U$.  
In the total intensity map, the western clump is the W3(OH) UC \HII\ region while the eastern clump is the W3(H$_2$O) hot core, where triangles mark the two 1.4~mm continuum peaks, A and C, from east to west.  
Two distinct and uniform distributions of polarized emission are observed in W3(H$_2$O) and W3(H$_2$O)-NW.  
Note that the total intensity in W3(OH) is complicated by the free-free continuum of the UC \HII\ region. 
% and is not normalized to dust polarized emission.  
Additional care is required to derive the polarization measurements.  
% A scale bar of 5\% polarization percentage is also shown.  
Contour levels correspond to $(-4, -2, 2, 4, 8, 12, 16, 20, 40, 60, 80) \times \sigma_I$.    
The angular resolution is 1\farcs5.  
\label{fpol}} %
\end{figure}

The $878 \; \mu\mathrm{m}$ polarization map is shown in Figure~\ref{fpol}.   
Line segments indicate the orientation of polarization angle, $\theta_\mathrm{obs}$, with their lengths proportional to the polarization percentage, $p$.     
Both the W3(H$_2$O) hot core and the W3(OH) UC \HII\ region are clearly detected in the Stokes $I$ map.  
Polarization is detected in W3(H$_2$O), W3(OH), and their vicinity, particularly in a faint, extended component in the north-west of W3(H$_2$O).  
In the W3(OH) UC \HII\ region, the total intensity is dominated by the free-free continuum, which is supposed to be unpolarized.   
Additional care is needed to estimate the relative contributions from dust and free-free emission to the total intensity in W3(OH), and this will be discussed elsewhere.   
%and will be discussed in Sect.~\ref{sec_w3oh}.  

%% 
% all the observational results, mainly P.A. rms
\subsection{Morphology and Properties of Detected Polarization} %  
In Table~\ref{tpol}, we group the detected polarization values based on their apparent association with distinct polarization subregions: W3(H$_2$O), W3(H$_2$O)-NW referring to the extended structure in the north-west of W3(H$_2$O), W3(H$_2$O)-E and W3(H$_2$O)-W referring to the two small regions in the east and west of W3(H$_2$O), respectively.  
With the exception of W3(H$_2$O)-NW, there is a general east-west orientation of $B_\bot$ across the whole region (Figure~\ref{fvla}).  

In W3(H$_2$O),  the polarization orientation is remarkably uniform, $\theta_\mathrm{obs} = -12^\circ - 20^\circ$.   
%with larger deviation in the two north-west segments, which may be affected by W3(H$_2$O)-NW with distinct polarization.  
The mean polarization angle weighted by observational uncertainty is $\langle \theta_\mathrm{obs} \rangle = 0^\circ \pm 2^\circ$.  
We compute the standard deviation, $\delta \theta_\mathrm{obs}$, with respect to $\langle \theta_\mathrm{obs} \rangle$ and obtain a value of $11.4^\circ$ (Table~\ref{tB}).  
This observed dispersion, $\delta \theta_\mathrm{obs}$, includes the measurement uncertainty, which is the root mean square (rms), $\delta \sigma_{\theta, \mathrm{obs}}$, of the observed uncertainties, $\sigma_{\theta, \mathrm{obs}}$.  
Hence, the intrinsic dispersion of polarization angle is given by $\delta \theta = (\delta \theta_\mathrm{obs}^2 - \delta \sigma_{\theta, \mathrm{obs}}^2)^{1/2} = 9.4^\circ$ and will be used to estimate the magnetic field strength of $B_\bot$ (Sect.~\ref{sec_jet}).   
In addition, the distribution of polarization angle in W3(H$_2$O)-NW (Figure~\ref{fpol}) also appears  fairly uniform.  
We compute for W3(H$_2$O)-NW a mean polarization angle of $\langle \theta_\mathrm{obs} \rangle = 54^\circ \pm 2^\circ$ and an intrinsic polarization angle dispersion of $\delta \theta = 10.2^\circ$.  

\subsection{The Magnetized Environment of the Synchrotron Jet in W3(H$_2$O) \label{sec_jet}} % 
\begin{figure} %
\epsscale{1.0} %
\plotone{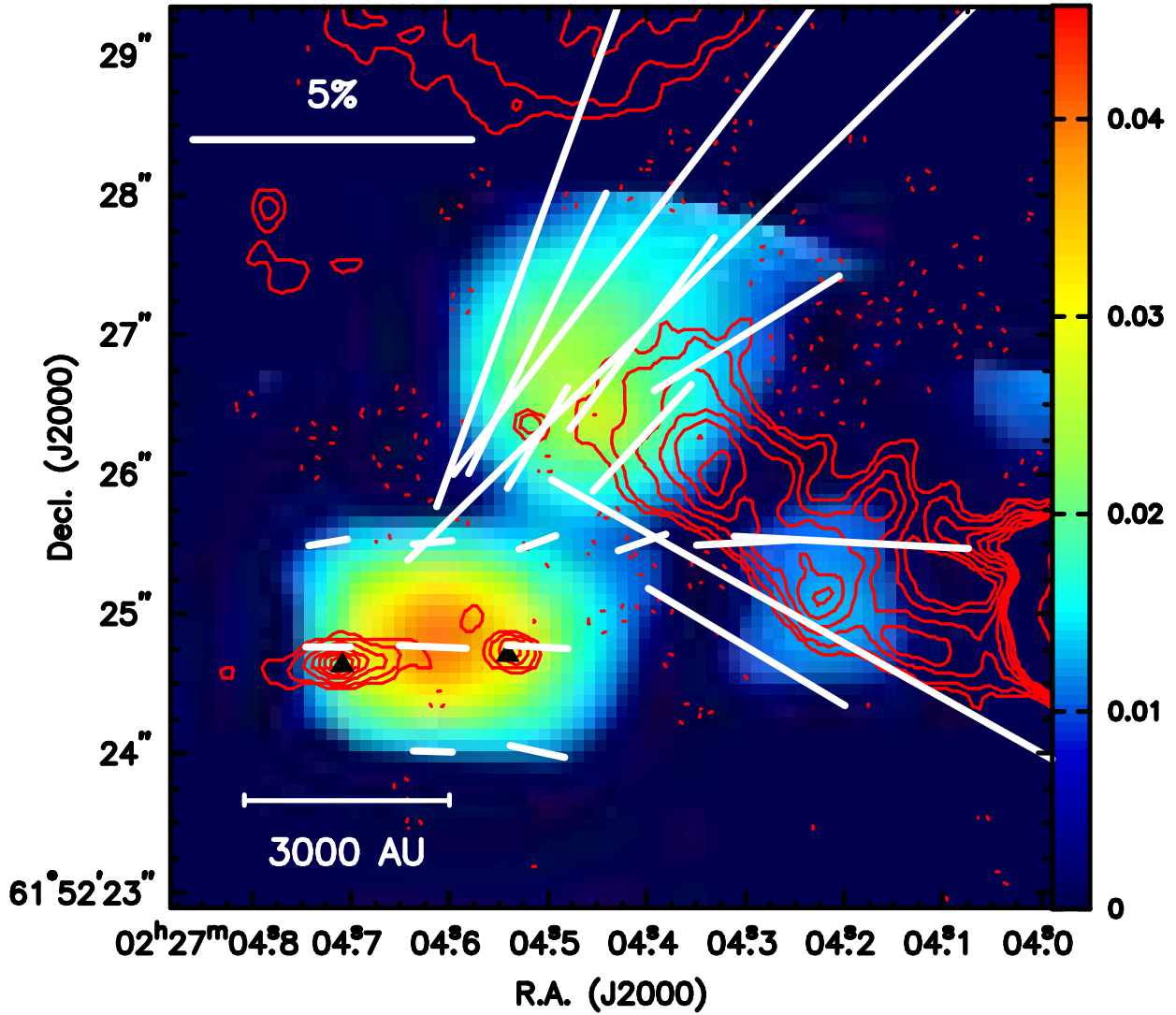} %  
\caption{The VLA 3.6~cm continuum map \markcite{Wilner1999ApJ513}(contours; Wilner {et~al.} 1999) overlaid on the $878 \; \mu\mathrm{m}$ linearly polarized intensity, $I_p$, map (colors).   
Line segments represent the inferred orientation of the projected magnetic fields, $B_\bot$.  
Triangles mark the two 1.4~mm continuum peaks in W3(H$_2$O), where the eastern peak, W3(H$_2$O)A, is associated with the synchrotron jet seen in the VLA $3.6 \; \mathrm{cm}$ map.  
%Red contour levels follow the levels used by \markcite{Wilner1999ApJ513}Wilner {et~al.} (1999).  
%The lowest contour levels are $30$ and $60 \; \mathrm{\mu Jy \, beam^{-1}}$, then the levels increase in steps of $60$ up to $360 \; \mathrm{\mu Jy \, beam^{-1}}$, followed by the four highest levels of 5, 10, 15, and $20 \; \mathrm{mJy \, beam^{-1}}$. 
Contour levels correspond to $30, 60, 120, 180, 240, 300, 360, 500 \; \mathrm{\mu Jy \, beam^{-1}}$.  
%In addition, the dust intensity map subtracted by the free-free continuum is also shown.  
\label{fvla} } %
\end{figure}

The synchrotron jet exhibits a double-sided morphology along the east-west direction centered at the millimeter continuum peak W3(H$_2$O)A in the VLA 3.6~cm map \markcite{Wilner1999ApJ513}(Figure~\ref{fvla}; Wilner {et~al.} 1999).  
Our uncertainty weighted mean polarization angle of $\langle \theta_\mathrm{obs} \rangle = 0^\circ \pm 2^\circ$ implies an east-west orientation of $B_\bot$, in good agreement with the direction of the synchrotron jet.   
In early VLA studies of the synchrotron jet \markcite{Reid:1995ib,Wilner1999ApJ513}(Reid {et~al.} 1995; Wilner {et~al.} 1999), an inhomogeneous  model of a  biconical geometry with opening angle of 0.1~rad successfully reproduced the observed properties.
The model gave a total magnetic field strength of $B(r) = 10 \; \mathrm{mG} \, (r/0\farcs2)^{-0.8}$, where $r$ is the angular distance from the center.    

The field strength of the projected magnetic field component, $B_\bot$, can be estimated if the dispersion of the polarization angle, $\delta \theta$, the mass density, $\rho$, and the velocity dispersion along line of sight, $\delta v_\mathrm{los}$, are measured.  
Considering a uniform magnetic field, \markcite{Chandrasekhar:1953ez}Chandrasekhar \& Fermi (1953; hereafter CF) first proposed a ``polarization-dispersion'' method to estimate $B_\bot$ based on transverse motions of Alfv\'{e}n waves, which are observable as dispersion in polarization angles, $\delta \theta$, from the mean orientation of the projected fields.  
If the velocity dispersion is isotropic and the transverse velocity dispersion is equal to $\delta v_\mathrm{los}$, one can obtain the plane-of-sky magnetic field strength  
\begin{eqnarray} %
B_\bot &=& Q \, \sqrt{4 \pi \rho} \, \frac{\delta v_\mathrm{los}}{\delta \theta} \nonumber \\
       &=& 68.6 \; \mathrm{mG} \left( \frac{Q}{0.5} \right) \left( \frac{\mu}{1.36} \right)^{1/2} \left( \frac{n_\mathrm{H_2}}{10^7 \; \mathrm{cm^{-3}}} \right)^{1/2} \left( \frac{\delta v_\mathrm{los}}{1 \; \mathrm{km \, s^{-1}}} \right) \left( \frac{\delta \theta}{1^\circ} \right)^{-1},   
\label{eqB} %
\end{eqnarray} % 
where $Q = 1$ corresponds to the original CF formula, and $\mu$ is the mean molecular weight. 
%This CF method have been applied in early studies of optical polarization maps to derive estimates of $B_\bot$ \markcite{Myers:1991ky}({Myers, P C} \& Goodman 1991).   
In a later study, \markcite{Ostriker:2001ky}Ostriker, Stone, \& Gammie (2001) performed 3D numerical MHD simulations with conditions in turbulent clouds and suggested a factor of $Q \sim 0.5$ applicable at varying observer orientations as long as $\delta \theta \lesssim 25^\circ$.   
\markcite{Chen2006ApJ639}Chen {et~al.} (2006) have studied the physical conditions in $\mathrm{W3(H_2O)}$ by analyzing the 1.4~mm and 2.8~mm continuum and the $\mathrm{CH_3CN}$ lines.  
%In the optically thin regime for the infrared photons where a temperature dependence of $r^{-0.4}$ is assumed, a density distribution of power-law index $p=1.52$ was found around both continuum peaks.   
Since our polarization observations do not resolve individual protostellar envelopes, we simply take their  mean value of $n_\mathrm{H_2} = 1.5 \times 10^7 \; \mathrm{cm^{-3}}$ and $\delta v_\mathrm{los} = 1.9 \; \mathrm{km \, s^{-1}}$ based on the $\mathrm{CH_3CN}$ spectra.    
Using Equation~(\ref{eqB}), we estimate the magnetic field strength to be $B_\bot = 17.0 \; \mathrm{mG}$.  
%\key{This estimate is an upper limit.}  
Since $\delta \theta_\mathrm{obs}$ can be lowered by the finite angular resolution and averaging along line of sight, this estimate of $B_\bot$ should be considered as an upper limit.   

When Zeeman measurements are available to provide the line-of-sight magnetic field strength, $B_\|$, it is possible to evaluate the total field strength with $B = (B_\bot^2 + B_\|^2)^{1/2}$.  
This offers an independent method to estimate the total field strength in comparison with the synchrotron model.  
The Zeeman measurements in $22 \; \mathrm{GHz}$ $\mathrm{H_2O}$ masers in W3(H$_2$O) yield a postshock field strength of $B_{\|,\mathrm{post}} = +42.1 \; \mathrm{mG}$ \markcite{Sarma2002ApJ580}(Sarma {et~al.} 2002).  
Depending on the degree of shock compression, the field could be amplified by a factor of about $20$ \markcite{Sarma2002ApJ580}(Sarma {et~al.} 2002) and renders an estimate for the preshock field strength of $B_\| = +2.1 \; \mathrm{mG}$.   
Hence, an estimate of the total field strength is about $B \simeq 17.1 \; \mathrm{mG}$, with orientation   close to the plane of sky.  
Though slightly higher, this total field strength is comparable to the value given by the synchrotron model \markcite{Reid:1995ib,Wilner1999ApJ513}(Reid {et~al.} 1995; Wilner {et~al.} 1999).  

Following \markcite{Crutcher:1999cd}Crutcher (1999), we estimate the observed-to-critical mass-to-magnetic flux ratio, $(M/\Phi_B)_\mathrm{obs/crit} = (N_\mathrm{H_2}/10^{24} \; \mathrm{cm^{-2}})/(B/10 \; \mathrm{mG}) = 0.54$, for W3(H$_2$O) with $N_\mathrm{H_2} = 9.2 \times 10^{23} \; \mathrm{cm^{-2}}$ \markcite{Chen2006ApJ639}(Chen {et~al.} 2006).  
This implies the core is magnetically subcritical and can be supported by the static magnetic field against the gravitational collapse.  
However, ongoing collapse in this region has been suggested by infall asymmetry in spectral profiles  \markcite{Wu:2003hw,Wu:2010dg}(Wu \& Evans 2003; Wu {et~al.} 2010).    
Since interferometric observations tend to underestimate column density and $B_\bot$ is an upper limit, our estimate of $(M/\Phi_B)_\mathrm{obs/crit}$ should be treated as a lower limit.  
Using the total field strength, the Alv\'{e}n velocity is given by $v_A = B/(4 \pi \rho)^{1/2} = 4.2 \; \mathrm{km \, s^{-1}} (B/10 \; \mathrm{mG})(n_\mathrm{H_2}/10^7 \; \mathrm{cm^{-3}})^{-1/2} = 5.9 \; \mathrm{km \, s^{-1}}$.   
The ratio of the turbulent to magnetic energy can be evaluated with $\beta_\mathrm{turb} = 3 (\delta v_\mathrm{los}/v_A)^2 = 0.31$, implying that the magnetic energy dominates over turbulence in this hot core \markcite{Girart:2009hi}(Girart {et~al.} 2009).  
%This is anticipated with the synchrotron jet as a display of magnetic activities.   

%% W3(H2O)-NW
Unlike the hot core, there are very limited observations addressing the extended structure W3(H$_2$O)-NW. 
We are not able to estimate the field strength for $B_\bot$ at the moment.    
The mean polarization angle of $\langle \theta_\mathrm{obs} \rangle = 54^\circ \pm 2^\circ$ in this extended structure appears quite different from the general north-south polarization orientation and implies $B_\bot$ to be at $\mathrm{P.A.} = -36^\circ$.
Based on its morphology and relative position with W3(OH) and the northern \HII\ region (Figure~\ref{fvla}), we speculate that W3(H$_2$O)-NW could be residual matter being pushed by the expansion of the northern \HII\ region and the champagne flow from W3(OH). 
%\key{Please check if you are comfortable with this speculation.}

%% Comparison with JCMT SCUBA results
\subsection{Comparison with Polarization Measurements on Large Scales} % 
\begin{figure} %
\epsscale{0.9} %
\plotone{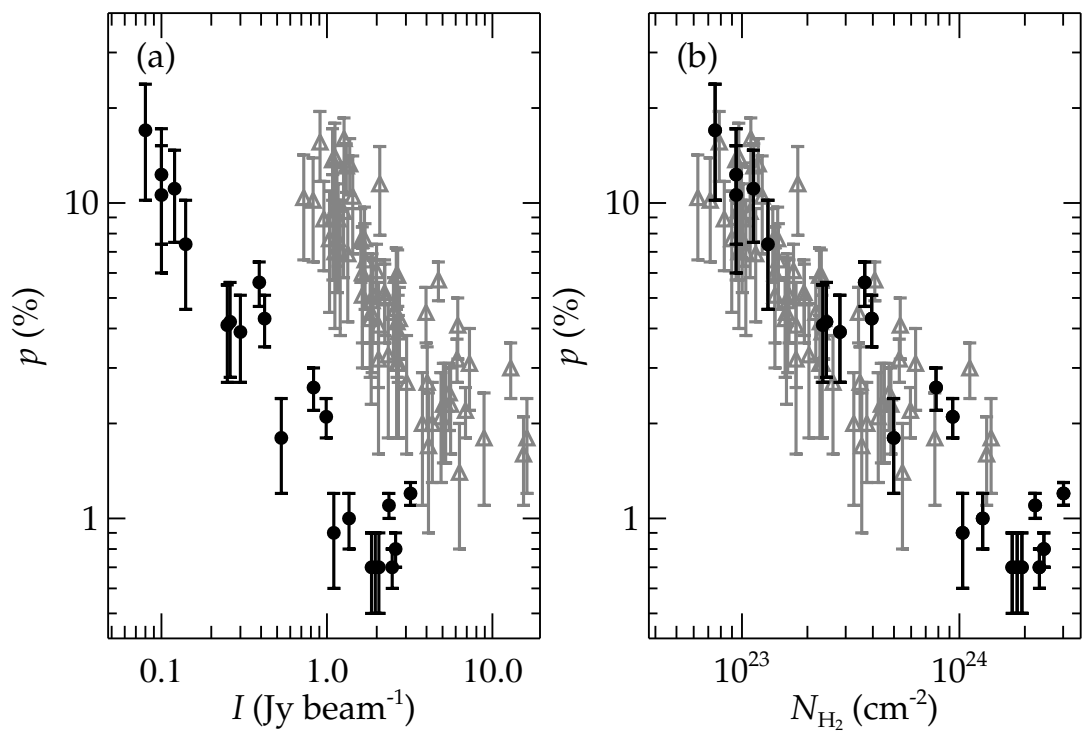} %
\caption{(a) Polarization percentage, $p$, vs. total intensity, $I$.  
Black dots are polarization measurements with the SMA whereas the gray triangles are measurements with the JCMT. 
(b) Polarization percentage, $p$, vs. beam-averaged column density, $N_\mathrm{H_2}$.  
Despite the great difference in angular resolutions and map extents between the SMA and JCMT  observations, a similar dependence is found with the beam averaged column density.   
%Black dots show  measurements in W3(H$_2$O), blue squares in W3(H$_2$O)-NW, red triangles in W3(H$_2$O)W, magenta crosses in W3(OH), and yellow diamand in W3(H$_2$O)-E.
\label{fpolm} } %
\end{figure}

Our polarization measurements show a decrease of polarization percentage, $p$, toward regions of higher intensity, $I$, (Figure~\ref{fpolm}a, black dots).  
To avoid complication caused by the UC \HII\ region, we exclude polarization measurements in W3(OH).      
Assuming a power-law dependence, we obtain a relationship of $\log p = (-0.73 \pm 0.05) \log I + (0.30 \pm 0.02)$ with a reduced chi-square of $\overline{\chi^2} = 2.2$.  
This ``depolarization'' phenomenon has been reported in many regions, e.g. L1755 \markcite{Lazarian:1997ht}(Lazarian {et~al.} 1997), Orion-KL \markcite{Rao:1998ew}(Rao {et~al.} 1998), OMC-3 \markcite{Matthews2001ApJ562}(Matthews, Wilson, \&  Fiege 2001), W51 \markcite{Lai2003ApJ598,Tang:2009ej}(Lai, Girart, \& Crutcher 2003; Tang {et~al.} 2009b), G5.89$-$0.39 \markcite{Tang:2009ek}(Tang {et~al.} 2009a).  
The reduction of polarization may be due to complexity of magnetic field geometry \markcite{Matthews2001ApJ562}(Matthews {et~al.} 2001), decreasing dust alignment efficiency \markcite{Lazarian:1997ht}(Lazarian {et~al.} 1997), beam smearing over small-scale structures \markcite{Rao:1998ew}(Rao {et~al.} 1998), and integration of varying magnetic domains along the line of sight.

%Comparing to the $850 \; \mu\mathrm{m}$ polarization measurements with the James Clerk Maxwell Telescope (JCMT), our SMA measurements nicely fill in the central  polarization ``hole'' in the JCMT polarization map \markcite{Matthews:2009cj}(Matthews {et~al.} 2009).    
The SMA polarization measurements nicely fill in the central polarization ``hole'' in the larger scale $850 \; \mu\mathrm{m}$ polarization map from the James Clerk Maxwell Telescope (JCMT) of Matthews {et~al.} (2009).
The three innermost JCMT measurements sit around the edge of the $37^{\prime\prime}$ SMA field of view.    
In addition, the JCMT measurements with an angular resolution of $20^{\prime\prime}$ also show the depolarization effect (Figure~\ref{fpolm}a; gray triangles).    
Despite the great difference in angular resolutions and map extents between the SMA and JCMT observations, the JCMT measurements yield a similar relationship of $\log p = (-0.72 \pm 0.05) \log I + (1.05 \pm 0.03)$ with $\overline{\chi^2} = 1.1$. 
%To understand this similarity, we convert the dust intensity to column density by assuming a dust opacity of $\kappa_\nu = 6 \times 10^{-3} \, (\lambda/1.22 \; \mathrm{mm})^{-1.5}$, a gas-to-dust mass ratio of $100$, a dust temperature of $20 \; \mathrm{K}$ \markcite{Wilson1993ApJ402}(Wilson, Gaume, \& Johnston 1993) for the JCMT measurements and $200 \; \mathrm{K}$ \markcite{Chen2006ApJ639}(Chen {et~al.} 2006) for the SMA measurements. 
To understand this similarity, we convert the dust intensity to column density with assumptions of a gas-to-dust mass ratio of $100$, a dust opacity of $\kappa_\nu = 10 \, (\lambda/250 \, \mu\mathrm{m})^{-\beta} \; \mathrm{cm^2 \, g^{-1}}$ \markcite{Hildebrand:1983tm}(Hildebrand 1983), and a dust temperature of $20$ and $105 \; \mathrm{K}$ for the JCMT and SMA measurements, respectively \markcite{Wilson1993ApJ402,Chen2006ApJ639,}(Wilson, Gaume, \& Johnston 1993; Wink {et~al.} 1994).  
For the central region, we derived $\beta = 0.9$ by comparing the flux density between $2.8 \; \mathrm{mm}$ (Chen {et~al.} 2006) and $878 \; \mu\mathrm{m}$ in maps made with visibilities of projected baselines within $8 - 207 \, \mathrm{k}\lambda$, the range that the two frequencies have in common.  
Such small $\beta$ is often attributed to grain growth in high-density environment (Miyake \& Nakagawa 1993) and is not valid in the outer region.  
To convert the JCMT measurements, we adopt $\beta = 1.6$ for active protostellar cores \markcite{Rathborne:2010bl}(Rathborne {et~al.} 2010). 
In Figure~\ref{fpolm}b, the polarization percentage shows a similar dependence with the beam averaged column density between the two populations of polarization measurements (Figure~\ref{fpolm}b).  
A fit using all the data gives $\log p = (-0.72 \pm 0.02) \log N_\mathrm{H_2} + (17.7 \pm 0.6)$ with $\overline{\chi^2} = 1.4$.
We suggest that the column density may be an important factor when interpreting the depolarization effects caused by decreasing alignment efficiency, beam smearing, and integration along line of sight.

%%%% Summary %%%%
\section{Summary} %
We present the first interferometric polarization map in the W3(OH) massive star-forming region observed with the SMA at $878 \; \mu\mathrm{m}$ with an angular resolution of 1\farcs5, corresponding to $3 \times 10^3 \; \mathrm{AU}$.    
The main findings are summarized as follows:
\begin{enumerate} %
\item The magnetic field orientation is well aligned with the synchrotron jet in W3(H$_2$O) and is close to the plane of sky with $B_\bot = 17.0 \; \mathrm{mG}$.    
The total field strength is estimated to be $17.1 \; \mathrm{mG}$, slightly higher but comparable to the estimate of $10 \; \mathrm{mG}$ from the synchrotron model.  
The magnetic energy dominates over turbulence in the hot core.  

\item The depolarization effect is discerned in both the SMA  and JCMT measurements.  
We find that the polarization percentage has a similar power-law dependence with the beam averaged column density for the two populations of polarization measurements.  
The column density may be an important factor to consider when interpreting the depolarization effect caused by decreasing alignment efficiency, beam smearing, and integration along line of sight.  

\item The extended emission structure in the north-west of W3(H$_2$O) has a significant polarization detection that suggests the $B_\bot$ orientation at $\mathrm{P.A.} = -36^\circ$.

\end{enumerate} %
 
%%%% Acknowledgments %%%%
\acknowledgments %
This research is supported by National Science Council of Taiwan through grant NSC 100-2112-M-007-004-MY2.  
Vivien Chen would like to thank Dr. B.~Matthews for the general guideline to convert JCMT SCUPOL flux density.  

%% example: {\it Facilities:} \facility{Nickel}, \facility{HST (STIS)}, \facility{CXO (ASIS)} %

%%%% Appendix %%%%
%\appendix %
%\section{} %

%%%% References %%%%
%\bibliographystyle{apj} %
%% \bibliography

%%%% Figures and Tables
%% Table 1 -- 
\begin{deluxetable}{cccr@{ $\pm$ }lr@{ $\pm$ }lc} %
\tablewidth{0pt} %
\tablecolumns{8} %
\tablecaption{Polarization measurements at $878 \; \mu\mathrm{m}$  \label{tpol}} %
\tablehead{ \colhead{R.A. Offset\tablenotemark{a}} & \colhead{Decl. Offset\tablenotemark{a}} & \colhead{$I$\tablenotemark{b}} & \multicolumn{2}{c}{$p$} & \multicolumn{2}{c}{$\theta_\mathrm{obs} \pm \sigma_{\theta,\mathrm{obs}}$} & \colhead{Remarks} \\
\colhead{(arcsec)} & \colhead{(arcsec)} & \colhead{$\mathrm{(Jy \, beam^{-1})}$}  & \multicolumn{2}{c}{(\%)} & \multicolumn{2}{c}{(deg)} & \colhead{}   
} %
\startdata %
\sidehead{$\mathrm{W3(H_2O)}$} \cline{1-2} % 
4.53 & -0.59 & 1.36 & 1.0 & 0.2 & -12 & 7 &  \\
5.28 & -0.59 & 1.96 & 0.7 & 0.2 & -2 & 7 &  \\
4.53 & 0.16 & 2.37 & 1.1 & 0.1 & -3 & 4 &  \\
5.28 & 0.16 & 3.20 & 1.2 & 0.1 & -2 & 2 & $I_p$ peak \\
6.03 & 0.16 & 2.60 & 0.8 & 0.1 & 0 & 5 &  \\
3.78 & 0.91 & 1.10 & 0.9 & 0.3 & 19 & 10 &  \\
4.53 & 0.91 & 2.06 & 0.7 & 0.2 & 20 & 7 &  \\
5.28 & 0.91 & 2.48 & 0.7 & 0.1 & 5 & 5 &  \\
6.03 & 0.91 & 1.86 & 0.7 & 0.2 & 10 & 7 &  \\ 
\sidehead{$\mathrm{W3(H_2O)}$-NW} \cline{1-2} %
3.78 & 1.66 & 0.83 & 2.6 & 0.4 & 47 & 4 &  \\
4.53 & 1.66 & 0.99 & 2.1 & 0.3 & 59 & 5 &  \\
3.03 & 2.41 & 0.30 & 3.9 & 1.2 & 32 & 8 &  \\
3.78 & 2.41 & 0.42 & 4.3 & 0.8 & 53 & 5 &  \\
4.53 & 2.41 & 0.39 & 5.6 & 0.9 & 64 & 4 &  \\
3.03 & 3.16 & 0.08 & 17.0 & 6.8 & 44 & 7 &  \\
3.78 & 3.16 & 0.12 & 11.1 & 3.6 & 53 & 7 &  \\
4.53 & 3.16 & 0.10 & 10.6 & 4.6 & 70 & 9 &  \\ 
\sidehead{W3(H$_2$O)-E and W3(H$_2$O)-W} \cline{1-2} %
9.78 & -0.59 & 0.14 & 7.4 & 2.8 & 23 & 9 & W3(H$_2$O)-E \\
2.28 & 0.16 & 0.10 & 12.3 & 4.9 & -29 & 8 &  \\
3.03 & 0.16 & 0.25 & 4.1 & 1.4 & -31 & 9 &  \\
2.28 & 0.91 & 0.26 & 4.2 & 1.4 & -3 & 9 &  \\
3.03 & 0.91 & 0.53 & 1.8 & 0.6 & 3 & 10 &  \\ \hline 
\sidehead{W3(OH) without free-free continuum subtraction} \cline{1-2} %
-0.72 & -1.34 & 0.38 & 2.5 & 0.9 & -75 & 10 &  \\
0.03 & -1.34 & 0.35 & 3.3 & 1.0 & -79 & 8 &  \\                    
-0.72 & -0.59 & 1.24 & 1.4 & 0.3 & -74 & 5 &  \\
0.03 & -0.59 & 1.27 & 1.5 & 0.3 & -73 & 5 &  \\                    
0.78 & -0.59 & 0.54 & 1.8 & 0.6 & -70 & 10 &  \\
-0.72 & 0.16 & 1.81 & 0.6 & 0.2 & -75 & 9 &  \\                    
0.03 & 0.16 & 1.98 & 0.7 & 0.2 & -68 & 7 &  \\                     
0.03 & 1.66 & 0.38 & 2.6 & 0.9 & 1 & 9 &  \\                       
0.78 & 1.66 & 0.23 & 4.2 & 1.5 & 9 & 10 &  \\
\enddata %
\tablenotetext{a}{Position offsets with respect to the map center at $(\alpha,\delta)$(J2000) = $\mathrm{(2^h27^m3.870^s,+61^\circ52^\prime24.60^{\prime\prime})}$. }
\tablenotetext{b}{The rms of the Stokes $I$ map is $0.027 \; \mathrm{Jy \, beam^{-1}}$.} 
\end{deluxetable} %

%% Table 2 -- 
\begin{deluxetable}{lccccccc} %
\tablewidth{0pt} %
\tablecolumns{8} %
\tablecaption{Dispersions of Polarization Angles \label{tB}} %
\tablehead{ \colhead{Region} & \colhead{$\langle \theta_\mathrm{obs} \rangle$} & \colhead{$\delta \theta_\mathrm{obs}$} & \colhead{$\delta \sigma_{\theta,\mathrm{obs}}$} & \colhead{$\delta \theta$} & \colhead{$n_\mathrm{H_2}$} & \colhead{$\delta v_\mathrm{los}$} & \colhead{$B_\bot$} \\
\colhead{} & \colhead{(deg)} & \colhead{(deg)} & \colhead{(deg)} & \colhead{(deg)} & \colhead{($\mathrm{10^7 \; cm^{-3}}$)} & \colhead{($\mathrm{km \, s^{-1}}$)} & \colhead{(mG)} 
} %
\startdata %
W3(H$_2$O)        & $0 \pm 2$ & 11.4 & 6.4 & 9.4 & 1.5 & 1.90 & 17.0 \\
W3(H$_2$O)-NW & $54 \pm 2$ & 12.0 & 6.4 & 10.2 & \dots & \dots & \dots \\
\enddata %
\end{deluxetable} %

%% Figure 1 -- Polarization maps

%% Figure 2 -- B orientation on VLA map

%% Figure 3 -- Polarized percentage vs Stoke I 

\end{document}